\documentclass[a4paper,11pt]{article}
\pdfoutput=1 

\usepackage{jcappub} 

\usepackage[T1]{fontenc} 
\usepackage{soul}

\title{\boldmath Measuring the cosmic homogeneity scale with SDSS-IV DR16 Quasars}

\author[a, b]{Rodrigo S. Gon\c{c}alves,}
\author[c]{Gabriela C. Carvalho,}
\author[a]{Uendert Andrade,}
\author[a]{Carlos A. P. Bengaly,}
\author[a]{Joel C. Carvalho}
\author[a]{and Jailson Alcaniz}

\affiliation[a]{Observat\'orio Nacional, 20921-400, Rio de Janeiro, RJ, Brasil}
\affiliation[b]{Departamento de F\'isica, Universidade Federal Rural do Rio de Janeiro, Serop\'edica, Rio de Janeiro 23897-000, Brazil}
\affiliation[c]{Universidade do Estado do Rio de Janeiro, Faculdade de Tecnologia, 27537-000, Resende - RJ, Brasil}

\emailAdd{rsousa@on.br}
\emailAdd{gabriela.coutinho@fat.uerj.br}
\emailAdd{uendertandrade@on.br}
\emailAdd{carlosbengaly@on.br}
\emailAdd{jcarvalho@on.br}
\emailAdd{alcaniz@on.br}

\abstract{We report measurements of the scale of cosmic homogeneity ($r_{h}$) using the recently released quasar sample of the sixteenth data release of the Sloan Digital Sky Survey (SDSS-IV DR16). We perform our analysis in 2 redshift bins lying in the redshift interval $2.2 < z < 3.2$ by means of the fractal dimension $D_2$. 
By adopting the usual assumption that $r_{h}$ is obtained when $D_2 \sim 2.97$, that is, within 1\% of $D_2=3$, we find the cosmic  homogeneity scale with a decreasing trend with redshift, and in good agreement with the $\Lambda$CDM prediction. Our results confirm the presence of a homogeneity scale in the spatial distribution of quasars as predicted by the fundamental assumptions of the standard cosmological model.}

\begin{document}
\maketitle
\flushbottom

\section{Introduction}
\label{sec:intro}

The advent of increasingly larger and high precision observational data sets has enabled us to test the standard model of Cosmology (SMC) with unprecedented precision, as recently done with Cosmic Microwave Background (CMB) observations~\cite{planck18}, Type Ia Supernova (SNIa) distances~\cite{pantheon18}, and large-scale clustering (LSS)~\cite{des19, sdss20, kids20}. Given that we are currently able to constrain the SMC parameters with nearly sub-percent precision, we should also be able to perform precise tests of the fundamental hypothesis underlying the SMC as well. One of these hypothesis corresponds to the Cosmological Principle (CP), which states that the Universe is statistically homogeneous and isotropic at large scales and described by the Friedmann-Lema\^itre-Robertson-Walker (FLRW) metric. Although it is a widely accepted assumption in modern Cosmology, it is yet to be proven through cosmological observations. Hence, it is paramount that we establish the CP as an observationally valid assumption, rather than a mathematical simplification of the actual Universe (see e.g.~\cite{Ellis:2008up} for a discussion).

Observational tests of cosmological isotropy have been carried out with null results for a significant deviation of such assumptions~\cite{Andrade:2018eta, Andrade:2019kvl, Bengaly:2019zhr, Akrami:2019bkn, Kazantzidis:2020tko, Migkas:2020fza, Secrest:2020has}. However, cosmic homogeneity can only be indirectly tested, as we observe down the past lightcone, not on time-constant hypersurfaces~\cite{Clarkson:2010uz, Maartens:2011yx, Clarkson:2012bg}. Still, we can test consistency between the FLRW hypothesis and cosmological observations, and thus confirm cosmic homogeneity given that we obtain evidence for both FLRW consistency tests and statistical isotropy. 

A possible method to probe consistency between observational data and the homogeneity assumption consists in performing counts-in-spheres of cosmic objects e.g. luminous red galaxies and quasars, so that we measure the scale where these counts are statistically similar to those obtained from random - thus homogeneous by construction up to Poisson noise -  catalogues. This will be hereafter referred as the {\emph {cosmic homogeneity scale}} ($r_h$), which denotes the transition of the locally lumpy Universe to a smoother, statistically homogeneous one. This test has been extensively carried out in the literature, and most authors confirmed the existence of such a scale~\cite{Hogg:2004vw, Yadav:2010cc, Scrimgeour:2012wt, Alonso:2014xca, Pandey:2015xea, Laurent:2016eqo, Ntelis:2017nrj, Goncalves:2017dzs, Goncalves:2018sxa, Pandey:2016, Sarkar:2009}. However, there are still claims of possible absence of such scale, as discussed for instance in~\cite{Labini:2009ke, Labini:2011dv, Park:2016xfp}, and also that these measurements could be biased by the survey window function~\cite{Heinesen:2020wai}. 

In this paper we estimate the cosmic homogeneity scale using the spatial distribution of quasar number counts from the recently released quasar sample (DR16Q) of the Sloan Digital Sky Survey IV (SDSS-IV)~\cite{DR16Q}. As in previous BOSS data releases, DR16 is divided into two different regions in the sky, named  south and north galactic cap and the latter was the region used in this paper. This data-set contains 120,000 quasars within the redshift interval $2.2 < z < 3.2$, so that it is possible to probe the scale of cosmic homogeneity at two redshift bins  by means of scaled counts-in-spheres. Our results are used to estimate a possible redshift dependence of quasar bias, and verify whether there is a transition from a lumpy to a homogeneous Universe as well as the scale at which such transition happens. Furthermore, we compare these results with the theoretical prediction of the standard scenario for the matter density perturbations. We find a good agreement with previous results using different types of tracers of large-scale structure. 

This paper is organised as follows. In Section 2 we briefly describe the observational data used in our analysis. Section 3 discusses the method and correlation function estimators adopted. The results and discussions are presented in Section 4. Section 5 summarises our main conclusions.

\section{Observational Data}
\label{sec:ObsData}

We perform our analysis using the latest release of data of the SDSS-IV. The DR16Q is currently the largest catalogue of spectroscopically selected quasars, as it is $99.8\%$ complete with few contamination ($0.3-1.3\%$). This catalogue contains 750,414 quasars, including 239,081 Ly$\alpha$ object corresponding a $25\%$ increase over DR12Q. The DR16Q catalogue has also been visually inspected, providing reliable classifications and redshifts~\cite{DR16Q}. Given the quality and redshift range of this data-set, the DR16Q allows us to expand the homogeneity scale measurements to higher redshifts, and test the redshift evolution of $r_h$ predicted by the SMC. 

We assume two redshift bins from DR16Q to provide two new homogeneity scales measurements considering quasars at $2.2<z<3.2$. The choice of the bins is based on the widest possible redshift bin that comprises the largest number of data points, and that provides the smallest possible relative error for the homogeneity scale as we increase the number of data points. Following this procedure, we end up with $\sim 19,000$ quasars in each redshift bin, which ranges from $2.2<z<2.4$, and from $2.5<z<3.2$ (see table 1). Moreover, we stress that there is no overlap between redshift bins to avoid any correlation that affects the two independently homogeneity scale estimates.  

\begin{table}[h!]
\centering
\begin{tabular}{|c|c|c|}
\hline
\,\,$z$ interval \,\,& \,\,$\bar{z}$ \,\,& \,\,$N_{\rm q}$\,\, \\
\hline
\,\,2.20-2.40 \,\,& \,\,2.30 \,\,&  19026 \\
\,\,2.50-3.20 \,\,& \,\,2.85 \,\,& 19093 \\
\hline
\end{tabular}
\caption{The two redshift bins used in the analysis and their properties: redshift range, mean redshift value, and number of quasars.}
\label{table1}
\end{table}

\section{Method}
\label{sec:Method}

\subsection{Fractal Correlation Dimension}
\label{subsec:FCD}

From the data described in the previous section, we identify the spatial three-dimension position of each quasar. This is carried out using the right ascension ($\alpha$), declination ($\delta$) and redshift ($z$) of each object. As we are exploring the three-dimensional approach to the homogeneity scale, we must convert the angles and redshifts into physical separation ($r$) between each pair of quasars $\{Q_i,Q_j\}$ according to
\begin{equation}\label{eq:r}
r=\sqrt{d(z_{\rm Q,1})^2 + d(z_{\rm Q,2})^2 - 2d(z_{\rm Q,1})d(z_{\rm Q,2})\cos{\theta}}\,;
\end{equation}
\begin{equation}\label{eq:costheta}
\cos{\theta} = \cos{[\sin{\delta_{\rm Q,1}}\sin{\delta_{\rm Q,2}} + \cos{\delta_{\rm Q,1}}\cos{\delta_{\rm Q,2}}\cos{(\alpha_{\rm Q,1} - \alpha_{\rm Q,2})}]} \,,
\end{equation}
where the radial comoving distance of each quasar $d(z_{\rm Q})$ reads
\begin{equation}
\label{eq:dz}
d(z_{\rm Q}) = \int_0^z \frac{cdz'}{H(z')}; \; H(z) = H_0\sqrt{\Omega_{\rm m}(1+z)^3 + \Omega_\Lambda}\,.  
\end{equation}
We assume a $\Lambda$CDM fiducial model with $\Omega_{\rm m} = 0.313$, $\Omega_{\Lambda} = 0.687$, $H_0 = 67.48 \, \mathrm{km/s/Mpc}$, which is consistent with the latest Planck release~\cite{planck18}. 

From these quantities, we can obtain the so-called scaled counts-in-sphere ($\cal N$), as well as the fractal dimension $D_2$~\cite{Scrimgeour:2012wt}. We obtain these quantities by comparing the observational data with random catalogues, and thus determine the scale where the former tends to the latter. To do so, we use the Peebles-Hauser (PH) correlation function, defining the PH estimator as~\cite{Laurent:2016eqo}. 
\begin{equation}\label{NPH}
{\cal N}(\!<\! r)\! \equiv \!
\frac{\sum^{r}_{\rho = 0} DD(\rho)}{\sum^{r}_{\rho = 0} RR(\rho)} \;,
\end{equation}
where $DD(r)$ is the pair of observed quasars counts within a separation radius $r$ normalised by the the total number of pairs, $N^{\rm obs}(N^{\rm obs} - 1)/2$. On the other hand $RR(r)$ is the equivalent counting for the random catalogues. Note that the estimator based on Landy-Szalay correlation function \cite{Laurent:2016eqo} could be used, but previous results found similar results with the PH estimator \cite{Ntelis:2017nrj, Goncalves:2018sxa}, hence we will just deploy the PH estimator this time around.

Note also that the estimators, ${\cal N}$ and $D_2$, are given in terms of the two-point correlation function (following previous works~\cite{Laurent:2016eqo, Ntelis:2017nrj, Goncalves:2018sxa}), instead of the number of points inside spheres with different radii, which would affect the results due to possible overlapping between spheres~\cite{Pandey:2013, Kraljic:2015}.

From $\cal N$, we can compute the fractal correlation dimension $D_2$, which are related by
\begin{equation}\label{defD2}
D^{\rm q}_{\rm 2}(r) \equiv {d\ln {\mathcal{N}(<r)} \over d\ln\!r} + 3 \;.
\end{equation}
As the distribution approaches the homogeneity scale, the scaled counts-in-sphere ${\cal N} \rightarrow 1$, and hence the fractal correlation dimension $D_2 \rightarrow 3$. In our analysis, we adopt the latter definition. The advantage of using it rather than ${\cal N}$ is that $D_2$ is not strongly correlated as $\mathcal{N}$ (see e.g. \cite{Ntelis:2017nrj} for a discussion). We define the the scale where $D^{\rm q}_{\rm 2}(r)$ reaches 1\% of the homogeneous definition, i.e., $D^{\rm q}_{\rm 2}(r) = 2.97$, as the cosmic homogeneity scale $r^{\rm q}_{h}$. As commonly accepted in the literature, this definition of threshold for the correlation dimension is due to the survey geometry and incompleteness~\cite{Scrimgeour:2012wt, Laurent:2016eqo, Ntelis:2017nrj}.


\subsection{Theoretical predictions and bias }
\label{subsec:th_pred}

We find the theoretical expectation for the cosmic homogeneity scale in a given range by means of the two-point correlation function from the matter density power spectrum 
\begin{equation}\label{TheoreticalPS}
\xi(s,\bar{z}) = \frac{1}{2\pi^2} \int P_m (\kappa,\bar{z})\frac{sin(\kappa s)}{\kappa s} \kappa^2 d\kappa \,,
\end{equation}
where the matter power spectrum is obtained from the {\sc CAMB} code~\cite{Lewis:1999bs} assuming the fiducial model previously described. Then we calculate the theoretical scaled count-in-spheres as
\begin{equation}\label{TheoreticalCursN}
{\cal N}_{\rm th} (< r,\bar{z}) = \frac{3}{4\pi r^3} \int_{0}^{r} (1+\xi(s,\bar{z})) 4 \pi s^2 ds \,,
\end{equation}
and the correspondent theoretical fractal correlation dimension according to
\begin{equation}\label{defD2m}
D^{\rm m}_{2}(r) \equiv {d\ln {\mathcal{N}_{\rm th} (<r)} \over d\ln\!r} + 3 \,.
\end{equation}

In order to compare the theoretically predicted homogeneity scale with the observed counterpart, we need to obtain how the quasars trace the underlying matter distribution, i.e., the quasar bias. This quantity can be computed directly from the correlation dimension, as given by~\cite{Laurent:2016eqo, Ntelis:2017nrj} 
\begin{equation}\label{RelationD2}
D^{\rm m}_{\rm 2}(r) = \frac{D^{\rm q}_{\rm 2}(r) - 3}{b^2} + 3 \,,
\end{equation}
where $b$ represents the quasar bias, which is fitted according to a specific model. 


\section{Analysis and Results}
\label{sec:Results}

\begin{figure}[t]
\centering
    \includegraphics[scale=0.47]{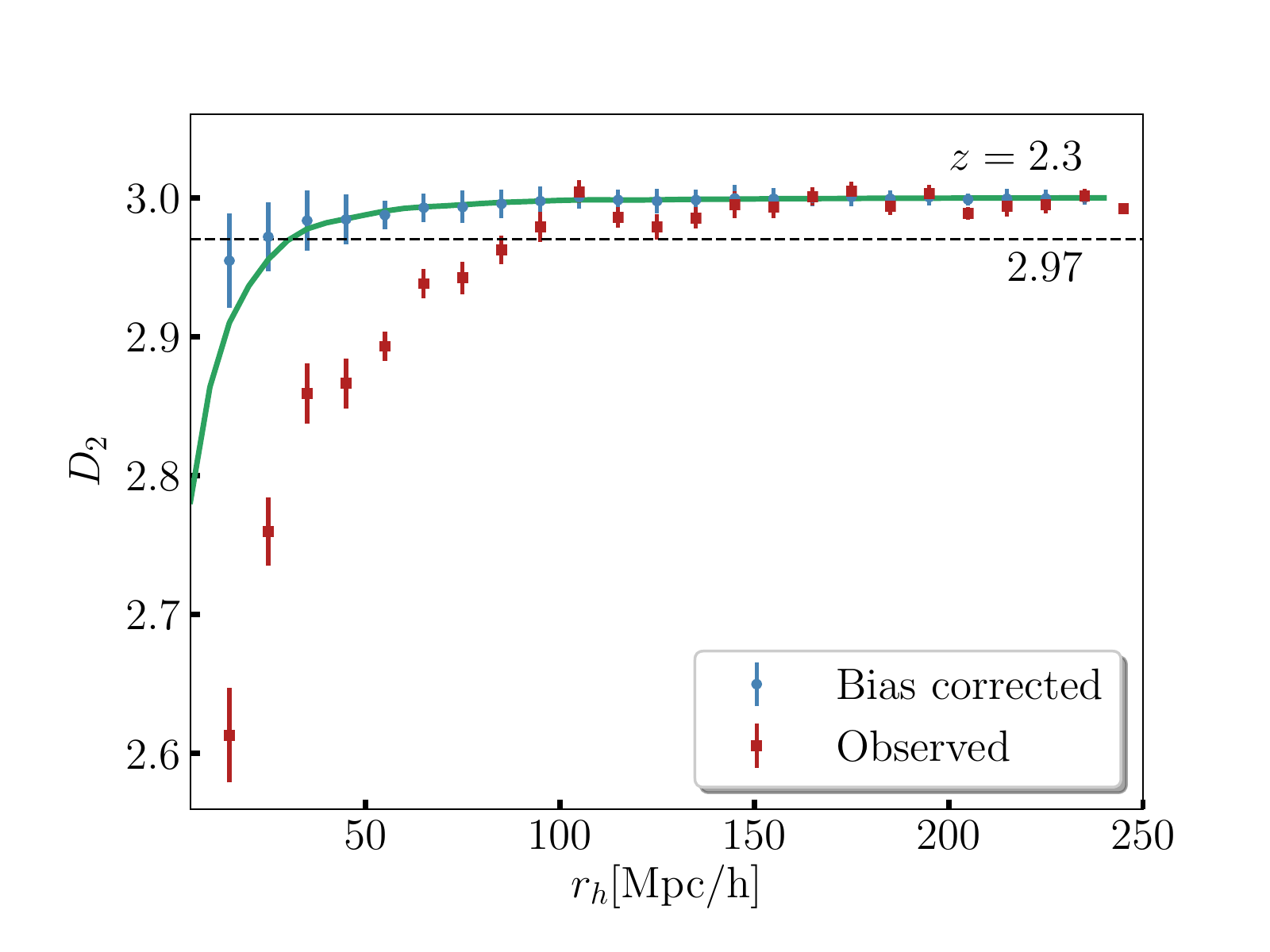}
    \includegraphics[scale=0.47]{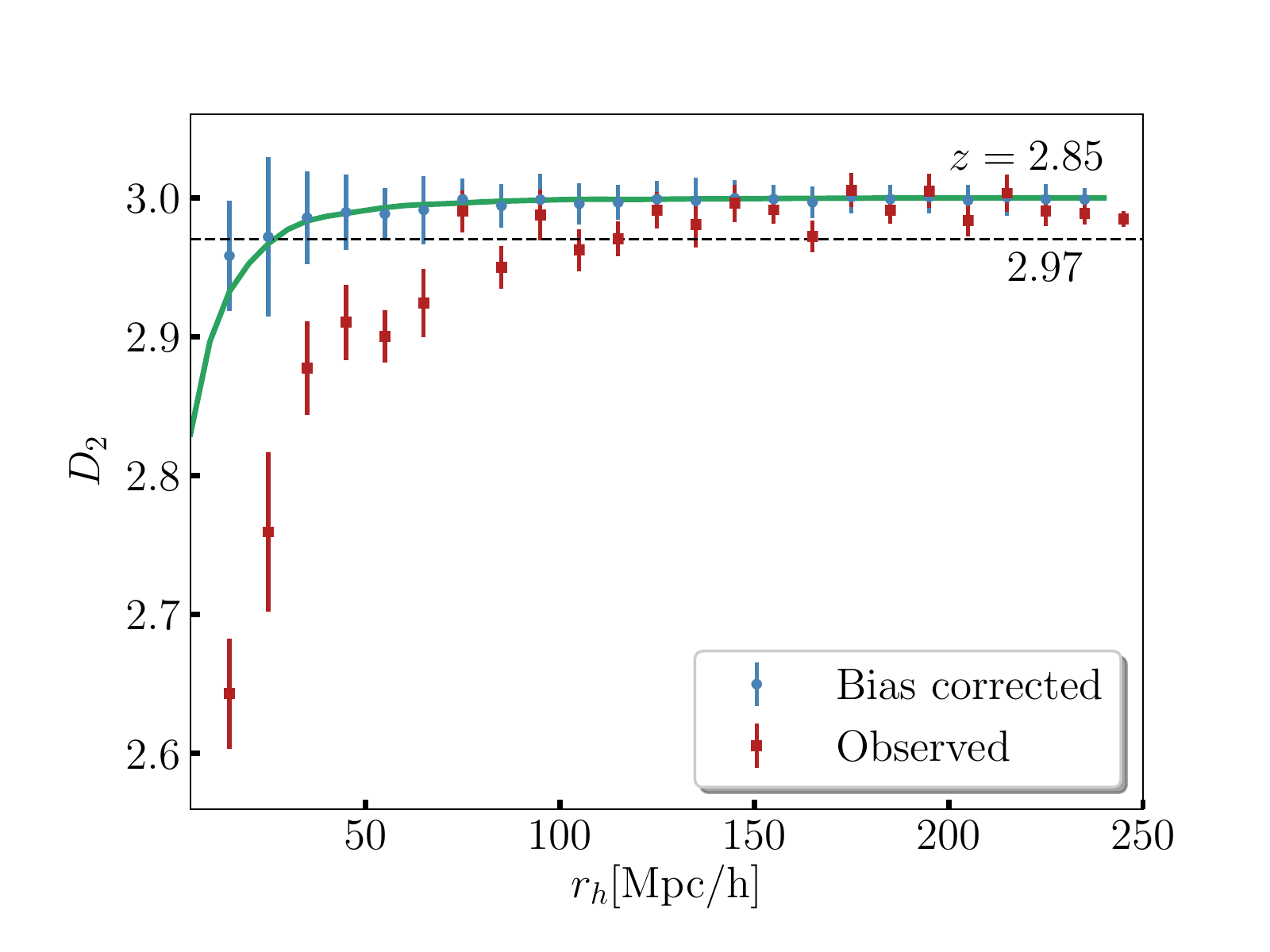}
    \caption{Fractal correlation dimension in each redshift bin. Blue points stand for observational values adjusted by the correspondent bias, green lines correspond to the theoretical prediction of the fiducial cosmological model, whereas the dotted blue line gives our homogeneity scale threshold, that is, $D^{\rm m}_{\rm 2} = 2.97$.}
    \label{D2eachZ} 
\end{figure}

In order to estimate the homogeneity scale in each redshift bin with the DR16Q data, firstly we obtain the correlation dimension as described in Eq.~\ref{defD2}. We use twenty random catalogues for this purpose, which implies that we have this number of $D^{\rm q}_{\rm 2}$. We fit a polynomial expression to describe the evolution of each $D^{\rm q}_{\rm 2}$, and then we find the corresponding homogeneity scale ($r^{\rm q}_{h}$) where $D^{\rm q}_{\rm 2} = 2.97$ for each realisation. Finally, we perform a bootstrap analysis over these twenty $r^{\rm q}_{h}$ values, obtaining their corresponding mean and standard deviation in each redshift bin. These results are shown in the second column in Table~\ref{Table:Rhom}. It is worth mentioning that in our analysis we follow~\cite{Goncalves:2018sxa, Laurent:2017} and adopt bootstrap instead of mock catalogues, as the former method seems to provide more conservative error estimates (see~\cite{Laurent:2017} for more details).

\begin{table}[!h]
\centering
\begin{tabular}{|c|c|c|c|}
\hline
$\bar{z}$ \,\,& \,\,$r^{\rm q}_{h}$\,\,& \,\,$r^{\rm m}_{h}$ \,\,& \,\, $r^{\rm th}_{h}$ \,\, \\
\hline
\,\,2.30 \,\,& 89.80 $\pm$ 5.05  \,\,&\,\, 29.66 $\pm$ 2.22 \,\,&\,\, 31.59 \\
\,\,2.85 \,\,& 71.35 $\pm$ 3.13  \,\,&\,\, 24.13 $\pm$ 3.49 \,\,&\,\, 27.56 \\
\hline
\end{tabular}
\caption{Homogeneity scale obtained from the estimator PH in each redshift slice. The first column provides the $r_{h}$ values obtained from the real data, the second column corresponds to the homogeneity scale after accounting for the bias as described in the text, $r^{\rm m}_{h}$, and the third column is the $r^{\rm th}_{h}$ estimated from the fiducial model following Eq.~\ref{defD2m}. All values are given in units of Mpc/h.}
\label{Table:Rhom}
\end{table}

We obtain the bias as described in Sec.~\ref{subsec:th_pred}. We assume a constant bias along each redshift bin ($b(r) = b_0$), and then we perform a $\chi^2$ analysis in Eq.~\ref{RelationD2}, whose best fit values for each redshift bin are $b(z = 2.3) = 2.53 \pm 0.03$ and $b(z = 2.85) = 2.92 \pm 0.08$, accordingly. Given these values, we obtain the correlation dimension for the matter distribution (blue points and bars in Figure~\ref{D2eachZ}), and hence the bias corrected values for the homogeneity scale $r^{\rm m}_{\rm h}$ - third column in Table~\ref{Table:Rhom}. 

It is worth mentioning that we explored the impact of different parameterisations of the bias on the radial comoving distance $d(z_{\rm Q})$, finding very similar results. In addition, we verified the possibility of a redshift evolution for the bias, finding a best-fitted parameterisation of $b(z) = b_1 + b_2(1+z)$, with $b_1 = 0.22$ and $b_2 = 0.7$  (see also ~\cite{Goncalves:2018sxa}). Still, our results are robust with respect to different bias choices. Moreover, we note that these results are in good agreement with previous estimates in the literature. For instance, $b(z= 0.53) = 1.38 \pm 0.10$~\cite{Shen:2013}, $b(z= 1) = 1.61 \pm 0.22$~\cite{Geach:2013}, $b(z= 1.41) = 1.92 \pm 0.50$~\cite{Ross:2009}, which shows that our results are compatible with the previous results within $2\sigma$ confidence level (CL).

We also compare the results measured from the real data, $r^{\rm m}_{h}$ with the theoretically predicted scale of homogeneity from Eq.~\ref{defD2m}, which is presented in the fourth column of~\ref{Table:Rhom}. We find that they are in excellent agreement with each other, showing that the observed homogeneity scale is consistent with the standard model.

\begin{figure}
    \centering
    \includegraphics[scale=0.5]{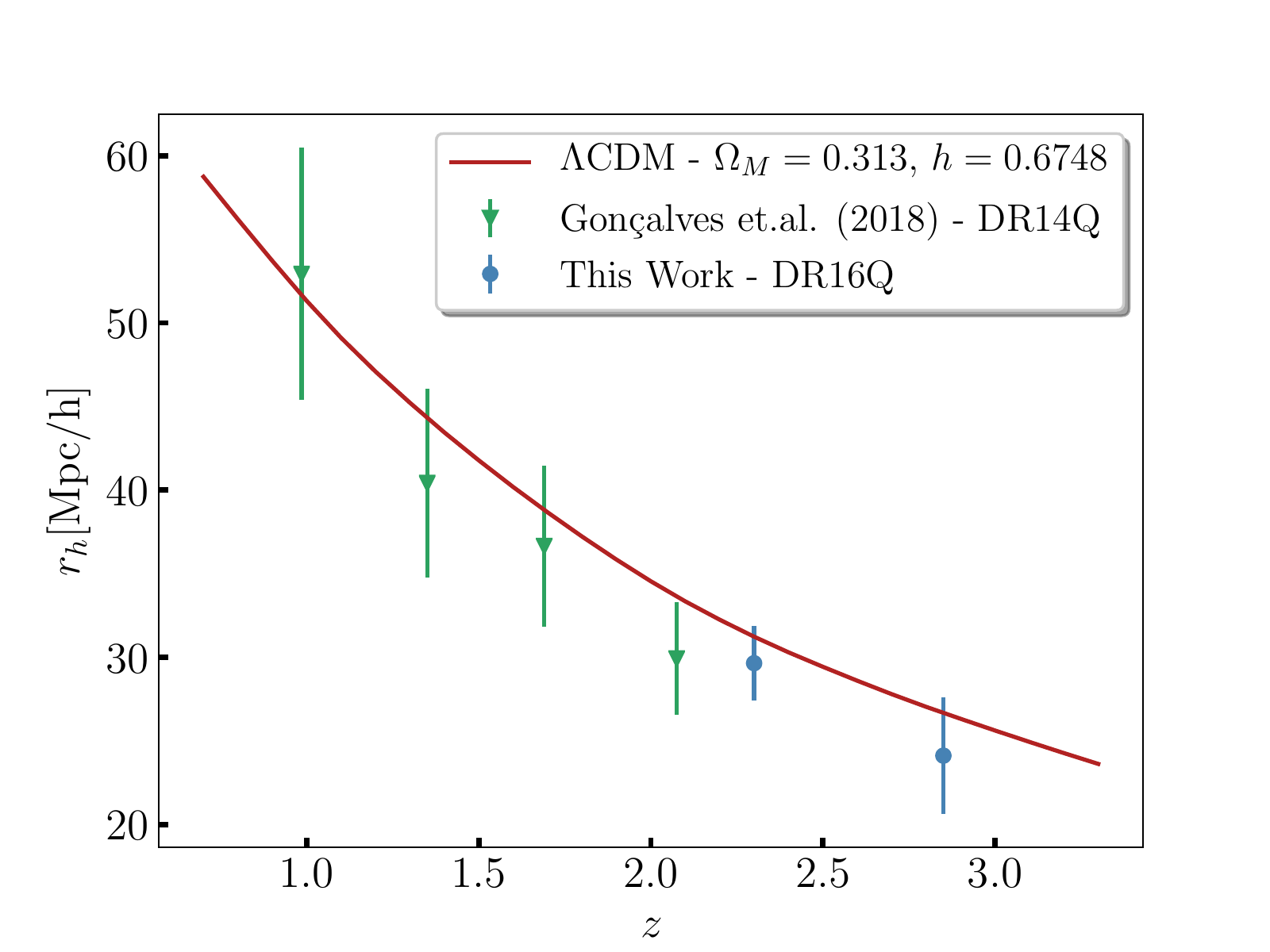}
    \caption{Homogeneity scale obtained in the present paper (blue points) and in~\cite{Goncalves:2018sxa}. The red line stands for the theoretical prediction for a  $\Lambda$CDM fiducial model ($\Omega_{\rm m} = 0.313$, $\Omega_{\Lambda} = 0.687$, $H_0 = 67.48 \, \mathrm{km/s/Mpc}$), the fiducial model used in the present paper.}
    \label{HomScale}
\end{figure}

For the sake of comparison, in Fig.~\ref{HomScale} (blue points) we plot the homogeneity scale for the matter distribution, as well as the values obtained in the previous work at  the corresponding redshift bins~\cite{Goncalves:2018sxa}. We also display the theoretical predictions from the fiducial model in this plot (red line), showing great agreement with our measurements - $1\sigma$ confidence level. These results confirm the decreasing trend of the homogeneity scale according to the redshift range, as obtained in previous analyses. 


\section{Conclusions}
\label{sec:Conclusions}

In this paper, we estimated the homogeneity scale using the latest quasar sample of the Sloan Digital Sky Survey, namely the SSS-IV DR16 data-set. To our knowledge, this is the first time that the transition to statistical homogeneity is assessed in such high redshift range ($2.2 < z < 3.2$). We split the original data set in two redshift slices, and verify whether there is indeed such characteristic scale, and also the presence of a possible evolution of $r_{h}^{\rm m}$ with respect to the redshift. Our analyses were performed using the scaled counts-in-spheres approach, $\mathcal{N}(<r)$, as well as the fractal correlation dimension $D_2$. We follow the usual definition that the cosmic homogeneity is attained where $D_2 \rightarrow 2.97$. We find that there is indeed evidence for such characteristic scale, i.e., $r^{\rm q}_h = 89.80 \pm 5.05 \; \mathrm{Mpc/h}$ ($\bar{z}=2.30$) and  $r^{\rm q}_h = 71.35 \pm 3.13 \; \mathrm{Mpc/h}$ ($\bar{z}=2.85$) at $1\sigma$ level.

In order to test consistency between data and the SMC, we fitted a constant bias in each redshift range, and thus obtained the scale of homogeneity of the matter distribution $r_{h}^{\rm m}$. We found that this scale indeed presents a decreasing trend with respect to the redshift, following previous results using the previous SDSS-IV DR14 quasar data. Moreover, we found that these results are in great agreement with the homogeneity scale predicted by the theoretical matter power spectrum, $r_{h}^{\rm th}$, in both redshift bins, showing consistency between our measurements and the standard model.  

Our results showed that there is a clear scale of cosmic homogeneity in the latest SDSS-IV quasar data, as predicted by the fundamental assumptions underlying the $\Lambda$CDM model. We plan to explore the interplay between the clustering bias and the homogeneity scale, in addition to other representations of the homogeneity scale rather than the fractal dimension $D_2$, in future works. We expect to improve our results, and thus establish the Cosmological Principle as a observationally valid hypothesis of the observed Universe, with the advent of next-generation surveys such as J-PAS~\cite{jpas,Bonoli:2020ciz}, DESI~\cite{desi}, Euclid~\cite{euclid}, and SKA~\cite{ska} 

\acknowledgments
We acknowledge the use of the Sloan Sky Digital Survey data. RSG is supported by CAPES through a PNPD fellowship. CB acknowledges financial support from the Programa de Capacita\c{c}\~ao Institucional PCI/ON. JSA is supported by Conselho Nacional de Desenvolvimento Cient\'{\i}fico e Tecnol\'ogico (CNPq 310790/2014-0) and Funda\c{c}\~ao de Amparo \`a Pesquisa do Estado do Rio de Janeiro (FAPERJ 233906). UA acknowledges financial support from CAPES and FAPERJ.


\begin{thebibliography}{99}

\bibitem{planck18} 
  N.~Aghanim {\it et al.} [Planck Collaboration],
   ``Planck 2018 results. VI. Cosmological parameters,''
  [arXiv:1807.06209].
 
\bibitem{pantheon18} 
  D.~M.~Scolnic {\it et al.},
  ``The Complete Light-curve Sample of Spectroscopically Confirmed SNe Ia from Pan-STARRS1 and Cosmological Constraints from the Combined Pantheon Sample,''
  Astrophys.\ J.\  {\bf 859}, no. 2, 101 (2018)
  [arXiv:1710.00845].

\bibitem{des19}
  T.~Abbott \textit{et al.} [DES],
  ``Dark Energy Survey Year 1 Results: Constraints on Extended Cosmological Models from Galaxy Clustering and Weak Lensing,''
  Phys. Rev. D \textbf{99} (2019) no.12, 123505
  [arXiv:1810.02499].
  
\bibitem{sdss20}
S.~Alam \textit{et al.} [eBOSS],
'`The Completed SDSS-IV extended Baryon Oscillation Spectroscopic Survey: Cosmological Implications from two Decades of Spectroscopic Surveys at the Apache Point observatory,''
[arXiv:2007.08991].

\bibitem{kids20}
C.~Heymans {\it et al.}
``KiDS-1000 Cosmology: Multi-probe weak gravitational lensing and spectroscopic galaxy clustering constraints,''
[arXiv:2007.15632].

\bibitem{Ellis:2008up}
G.~F.~R.~Ellis,
``Dark matter and dark energy proposals: maintaining cosmology as a true science?,''
EAS Publ. Ser. \textbf{36} (2009), 325-336
[arXiv:0811.3529].

\bibitem{Andrade:2018eta}
U.~Andrade, C.~A.~P.~Bengaly, B.~Santos and J.~S.~Alcaniz,
``A Model-independent Test of Cosmic Isotropy with Low-z Pantheon Supernovae,''
Astrophys. J. \textbf{865} (2018) no.2, 119
[arXiv:1806.06990].

\bibitem{Andrade:2019kvl}
U.~Andrade, C.~A.~P.~Bengaly, J.~S.~Alcaniz and S.~Capozziello,
``Revisiting the statistical isotropy of GRB sky distribution,''
Mon. Not. Roy. Astron. Soc. \textbf{490} (2019) no.4, 4481-4488
[arXiv:1905.08864].

\bibitem{Bengaly:2019zhr}
C.~A.~P.~Bengaly, R.~Maartens, N.~Randriamiarinarivo and A.~Baloyi,
``Testing the Cosmological Principle in the radio sky,''
JCAP \textbf{09} (2019), 025
[arXiv:1905.12378].

\bibitem{Akrami:2019bkn}
Y.~Akrami \textit{et al.} [Planck],
``Planck 2018 results. VII. Isotropy and Statistics of the CMB,''
[arXiv:1906.02552].

\bibitem{Kazantzidis:2020tko}
L.~Kazantzidis and L.~Perivolaropoulos,
``Hints of a Local Matter Underdensity or Modified Gravity in the Low $z$ Pantheon data,''
Phys. Rev. D \textbf{102} (2020) no.2, 023520
[arXiv:2004.02155].

\bibitem{Migkas:2020fza}
K.~Migkas, G.~Schellenberger, T.~H.~Reiprich, F.~Pacaud, M.~E.~Ramos-Ceja and L.~Lovisari,
``Probing cosmic isotropy with a new X-ray galaxy cluster sample through the $L_{\text{X}}-T$ scaling relation,''
Astron. Astrophys. \textbf{636} (2020), A15
[arXiv:2004.03305].

\bibitem{Secrest:2020has}
N.~Secrest, S.~von Hausegger, M.~Rameez, R.~Mohayaee, S.~Sarkar and J.~Colin,
``A Test of the Cosmological Principle with Quasars,''
[arXiv:2009.14826].

\bibitem{Clarkson:2010uz}
C.~Clarkson and R.~Maartens,
``Inhomogeneity and the foundations of concordance cosmology,''
Class. Quant. Grav. \textbf{27} (2010), 124008
[arXiv:1005.2165].

\bibitem{Maartens:2011yx}
R.~Maartens,
``Is the Universe homogeneous?,''
Phil. Trans. Roy. Soc. Lond. A \textbf{369} (2011), 5115-5137
[arXiv:1104.1300].

\bibitem{Clarkson:2012bg}
C.~Clarkson,
``Establishing homogeneity of the universe in the shadow of dark energy,''
Comptes Rendus Physique \textbf{13} (2012), 682-718
[arXiv:1204.5505].

\bibitem{Hogg:2004vw}
D.~W.~Hogg, D.~J.~Eisenstein, M.~R.~Blanton, N.~A.~Bahcall, J.~Brinkmann, J.~E.~Gunn and D.~P.~Schneider,
``Cosmic homogeneity demonstrated with luminous red galaxies,''
Astrophys. J. \textbf{624} (2005), 54-58
[arXiv:astro-ph/0411197].

\bibitem{Sarkar:2009}
P.~Sarkar, J.~Yadav, B.~Pandey, S.~Bharadwaj
``The  scale  of  homogeneity  of  the  galaxy  distribution  in  SDSS  DR6,''
Mon. Not. Roy. Astron. Soc. \textbf{399} (2009) L128
[arXiv:0906.3431].

\bibitem{Yadav:2010cc}
J.~K.~Yadav, J.~S.~Bagla and N.~Khandai,
``Fractal Dimension as a measure of the scale of Homogeneity,''
Mon. Not. Roy. Astron. Soc. \textbf{405} (2010), 2009
[arXiv:1001.0617].

\bibitem{Scrimgeour:2012wt}
M.~Scrimgeour {\it et al.}
``The WiggleZ Dark Energy Survey: the transition to large-scale cosmic homogeneity,''
Mon. Not. Roy. Astron. Soc. \textbf{425} (2012), 116-134
[arXiv:1205.6812].

\bibitem{Alonso:2014xca}
D.~Alonso, A.~I.~Salvador, F.~J.~Sánchez, M.~Bilicki, J.~García-Bellido and E.~Sánchez,
``Homogeneity and isotropy in the Two Micron All Sky Survey Photometric Redshift catalogue,''
Mon. Not. Roy. Astron. Soc. \textbf{449} (2015) no.1, 670-684
[arXiv:1412.5151].

\bibitem{Pandey:2015xea}
B.~Pandey and S.~Sarkar,
``Testing homogeneity in the Sloan Digital Sky Survey Data Release Twelve with Shannon entropy,''
Mon. Not. Roy. Astron. Soc. \textbf{454} (2015) no.3, 2647-2656
[arXiv:1507.03124].

\bibitem{Laurent:2016eqo}
P.~Laurent {\it et al.}
``A 14 $h^{-3}$ Gpc$^3$ study of cosmic homogeneity using BOSS DR12 quasar sample,''
JCAP \textbf{11} (2016), 060
[arXiv:1602.09010].

\bibitem{Ntelis:2017nrj}
P.~Ntelis {\it et al.}
``Exploring cosmic homogeneity with the BOSS DR12 galaxy sample,''
JCAP \textbf{06} (2017), 019
[arXiv:1702.02159].

\bibitem{Goncalves:2017dzs}
R.~S.~Gonçalves, G.~C.~Carvalho, C.~A.~P.~Bengaly {\it et al.}
``Cosmic homogeneity: a spectroscopic and model-independent measurement,''
Mon. Not. Roy. Astron. Soc. \textbf{475} (2018) no.1, L20-L24
[arXiv:1710.02496].

\bibitem{Goncalves:2018sxa}
R.~S.~Gonçalves, G.~C.~Carvalho, C.~A.~P.~Bengaly, J.~C.~Carvalho and J.~S.~Alcaniz,
``Measuring the scale of cosmic homogeneity with SDSS-IV DR14 quasars,''
Mon. Not. Roy. Astron. Soc. \textbf{481} (2018) no.4, 5270-5274
[arXiv:1809.11125].

\bibitem{Pandey:2016}
S.~Sarkar and B.~Pandey,
``An  information  theory  based  search  for  homogeneity  on  the  largest accessible scale,''
Mon. Not. Roy. Astron. Soc. \textbf{463} (2016) L12
[arXiv:1607.06194].


\bibitem{Labini:2009ke}
F.~S.~Labini, N.~L.~Vasilyev, Y.~V.~Baryshev and M.~Lopez-Corredoira,
``Absence of anti-correlations and of baryon acoustic oscillations in the galaxy correlation function from the Sloan Digital Sky Survey DR7,''
Astron. Astrophys. \textbf{505} (2009), 981-990
[arXiv:0903.0950].

\bibitem{Labini:2011dv}
F.~S.~Labini,
``Very large scale correlations in the galaxy distribution,''
EPL \textbf{96} (2011) no.5, 59001
[arXiv:1110.4041].

\bibitem{Park:2016xfp}
C.~G.~Park, H.~Hyun, H.~Noh and J.~c.~Hwang,
``The cosmological principle is not in the sky,''
Mon. Not. Roy. Astron. Soc. \textbf{469} (2017) no.2, 1924-1931
[arXiv:1611.02139].

\bibitem{Heinesen:2020wai}
A.~Heinesen,
``Cosmological homogeneity scale estimates are dressed,''
[arXiv:2006.15022].

\bibitem{DR16Q}
B. W.~Lyke  {\it et al.},``The Sloan Digital Sky Survey Quasar Catalog: Sixteenth Data Release", 
[arXiv:2007.09001]. 

\bibitem{Lewis:1999bs}
A.~Lewis, A.~Challinor and A.~Lasenby,
``Efficient computation of CMB anisotropies in closed FRW models,''
Astrophys. J. \textbf{538} (2000), 473-476
[arXiv:astro-ph/9911177].

\bibitem{Pandey:2013}
B.~Pandey,
``A method for testing the cosmic homogeneity with Shannon entropy,''
Mon. Not. Roy. Astron. Soc. \textbf{430} (2013) no.4, 3376–3382
[arXiv:1301.4961].

\bibitem{Kraljic:2015}
D.~Kraljic,
``Characterizing cosmic inhomogeneity with anomalous diffusion,''
Mon. Not. Roy. Astron. Soc. \textbf{451} (2015) no.4, 3393–3399
[arXiv:1410.4107].

\bibitem{Laurent:2017}
P.~Laurent {\it et al.}
``Clustering of quasars in SDSS-IV eBOSS : study of potential systematics and bias determination,''
JCAP \textbf{07} (2017), 017
[arXiv:1602.09010].


\bibitem{Shen:2013}
Y.~Shen  {\it et al.},
``Cross-Correlation of SDSS DR7 Quasars and DR10 BOSS Galaxies: The Weak Luminosity Dependence of Quasar Clustering at $z \sim 0.5$,''
Astrophys. J. \textbf{778} (2013), 98
[arXiv:1212.4526v1].

\bibitem{Geach:2013}
J.E..~Geach  {\it et al.},
``A Direct Measurement of the Linear Bias of Mid-infrared-selected Quasars at $z \approx 1$ Using Cosmic Microwave Background Lensing,''
Astrophys. J. Let. \textbf{776} (2013), L41
[arXiv:1307.1706]

\bibitem{Ross:2009}
N.P..~Ross  {\it et al.},
``Clustering of Low-Redshift (z <= 2.2) Quasars from the Sloan Digital Sky Survey,''
Astrophys. J. Let. \textbf{697} (2009), 1634
[arXiv:0903.3230]

\bibitem{jpas}
N.~Benitez \textit{et al.} [J-PAS],
``J-PAS: The Javalambre-Physics of the Accelerated Universe Astrophysical Survey,''
[arXiv:1403.5237].

\bibitem{Bonoli:2020ciz}
S.~Bonoli \textit{et al.} [J-PAS]
``The miniJPAS survey: a preview of the Universe in 56 colours,''
[arXiv:2007.01910].

\bibitem{desi}
A.~Aghamousa \textit{et al.} [DESI],
``The DESI Experiment Part I: Science,Targeting, and Survey Design,''
[arXiv:1611.00036].

\bibitem{ska}
D.~J.~Bacon \textit{et al.} [SKA],
``Cosmology with Phase 1 of the Square Kilometre Array: Red Book 2018: Technical specifications and performance forecasts,''
Publ. Astron. Soc. Austral. \textbf{37} (2020), e007
[arXiv:1811.02743].

\bibitem{euclid}
A.~Blanchard \textit{et al.} [Euclid],
``Euclid preparation: VII. Forecast validation for Euclid cosmological probes,''
[arXiv:1910.09273].

\end{thebibliography}
\end{document}